\documentclass[conference]{IEEEtran}
\IEEEoverridecommandlockouts
\usepackage{amsmath,amssymb,amsfonts}
\usepackage{algorithmic}
\usepackage{graphicx}
\usepackage{textcomp}
\usepackage{fancyhdr}
\usepackage{gensymb}
\usepackage{array}
\usepackage{makecell}
\usepackage{xcolor}
\usepackage{comment}
\usepackage{rotating}
\usepackage{afterpage}
\usepackage{tabularx}
\usepackage{diagbox}
\usepackage{array, caption, hhline}
\usepackage{enumitem} 
\usepackage{booktabs}
\usepackage{lipsum}

\def\BibTeX{{\rm B\kern-.05em{\sc i\kern-.025em b}\kern-.08em
    T\kern-.1667em\lower.7ex\hbox{E}\kern-.125emX}}
\usepackage[backend=biber,sorting=none,doi=false,isbn=false,url=false,eprint=false,style=ieee]{biblatex}
\begin{document}
\title{US Microelectronics Packaging Ecosystem: Challenges and Opportunities\\
}

\author{
    \IEEEauthorblockN{Rouhan Noor\IEEEauthorrefmark{1}, Himanandhan Reddy Kottur\IEEEauthorrefmark{1}, Patrick J Craig\IEEEauthorrefmark{1}, Liton Kumar Biswas\IEEEauthorrefmark{1}, M Shafkat M Khan\IEEEauthorrefmark{1},\\Nitin Varshney\IEEEauthorrefmark{1}, Hamed Dalir\IEEEauthorrefmark{1}, Elif Ak\c{c}al{\i}\IEEEauthorrefmark{1}, Bahareh Ghane Motlagh\IEEEauthorrefmark{2}, Charles Woychik\IEEEauthorrefmark{3},\\ Yong-Kyu Yoon\IEEEauthorrefmark{1} and Navid Asadizanjani\IEEEauthorrefmark{1}}
    \IEEEauthorblockA{\IEEEauthorrefmark{1}University of Florida, Gainesville, FL, USA\\ \IEEEauthorrefmark{2}BioNIUM, University of Miami, Coral Gables, FL, USA \\ \IEEEauthorrefmark{3}Skywater Technology Foundry, Bloomington, USA}
    \IEEEauthorblockA{Corresponding emails: noorrouhan@ufl.edu, nasadi@ufl.edu}
}

\maketitle

\begin{abstract}
The semiconductor industry is experiencing a significant shift from traditional methods of shrinking devices and reducing costs. Chip designers actively seek new technological solutions to enhance cost-effectiveness while incorporating more features into the silicon footprint. One promising approach is Heterogeneous Integration (HI), which involves advanced packaging techniques to integrate independently designed and manufactured components using the most suitable process technology. However, adopting HI introduces design and security challenges. To enable HI, research and development of advanced packaging is crucial. The existing research raises the possible security threats in the advanced packaging supply chain, as most of the Outsourced Semiconductor Assembly and Test (OSAT) facilities/vendors are offshore. To deal with the increasing demand for semiconductors and to ensure a secure semiconductor supply chain, there are sizable efforts from the United States (US) government to bring semiconductor fabrication facilities onshore. However, the US-based advanced packaging capabilities must also be ramped up to fully realize the vision of establishing a secure, efficient, resilient semiconductor supply chain.  Our effort was motivated to identify the possible bottlenecks and weak links in the advanced packaging supply chain based in the US. 
\end{abstract}


\section{Introduction}
The pervasive presence of electronic devices is profoundly transforming our way of life and work, becoming deeply embedded in our daily routines. In today's digital-driven economy, the widespread use of high-speed devices and seamless connectivity generates an immense volume of data. Various critical systems, including autonomous cars, data centers, and Artificial Intelligence (AI) systems, rely on capturing, storing, and analyzing this big data to enable data-driven transactions. Integrated Circuits (ICs) play a pivotal role in supporting the evolution of data processing, high-performance computing, and wireless communication. Today's cutting-edge ICs provide high-bandwidth memory, multiple processing cores, and high-speed input/output (I/O) ports. The presence of these cutting-edge ICs can be primarily credited to Moore's Law, which has driven the semiconductor sector to consistently produce ICs that are faster, smaller, and more cost-effective. However, this long-standing law faces limitations due to rising fabrication costs, power dissipation challenges, and yield issues associated with advanced technology nodes.

In response to the hurdles encountered in the traditional scaling of CMOS technology, the International Technology Roadmap for Semiconductors (ITRS) 2015 introduced a forward-looking strategy to sustain the historical advancement of CMOS technology\cite{dcadmin_2015_2015}. This vision centers around adopting HI as a viable solution. HI involves integrating individually designed and fabricated components on a substrate layer known as an interposer, allowing them to function like a System on Chip (SoC) collectively.
\begin{figure}[htp]
    \centering
    \includegraphics[width=8.5cm]{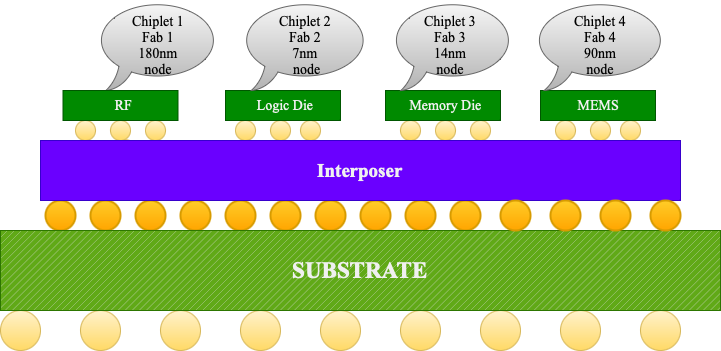}
    \caption{System-in-Package (SiP) for HI}
    \label{fig:galaxy}
\end{figure}

HI amalgamates separately manufactured components with varying technology nodes and functionalities, resulting in a more advanced assembly known as a Multi-Chip Module (MCM) or System in Package (SiP) (Fig. 1). SiPs offer expanded functionality and improved operational characteristics that are difficult to accomplish with a single-die SoC approach. Diverse components, such as chiplets, active/passive parts, and MEMS devices, can be integrated as a unified package into the SiP. Chiplets, for example, are individually fabricated silicon dies \cite{10239134} designed explicitly for targeted functions like memory, analog-mixed signal processing, radio frequency (RF), or processors. The SiP can be integrated through adjacent placement (2.5D) or vertical stacking (3D) of chiplets on the interposer.

While HI offers numerous advantages, further research and development are needed to enhance its effectiveness. To realize the vision of HI and continue Moore's Law, packaging technology needs to be improved to standardize interconnecting interfaces and communication protocols and ensure secure design. For instance, packaging methods should more efficiently utilize space to accommodate smaller form factors. That's why the traditional legacy packaging technology is insufficient to continue the HI and Moore's law. Moreover, the design of interconnecting interfaces for chiplets must meet specific criteria for speed, power, and mitigating crosstalk issues. The influence of high-performance computing, 5G, and AI leads to amplified semiconductor speed, heightened interconnect density, reduced pad pitch, expanded chip size, and increased power dissipation. There is a need to assemble and stack chips and dies vertically to enable further scaling with faster interconnection, such as developing 2D, 2.5D, and 3D stacking. These factors pose challenges to continuing with traditional packaging technology, and significant research is ongoing to develop this advanced packaging to cope with the requirements to realize HI. 

Many key players in the semiconductor sector, including integrated device manufacturers (IDM) like Intel, Micron, and Samsung, fabless design firms such as IBM and AMD, foundries like TSMC and Samsung, as well as OSATs like Amkor and TSMC, are deeply involved in the advancement of HI solutions. For instance, AMD EPYC and Intel Lakefield processors are commercially available examples of 3D SiPs\cite{updated_amd_2020}\cite{solution_video_nodate}. The Defense Advanced Research Projects Agency (DARPA) shares a similar vision through its Common Heterogeneous Integration and IP Reuse Strategies (CHIPS) Program, which aims to advance trusted microelectronics for the applications and technology requirements of the US Department of Defense (DoD)\cite{ravi_chips_2021}\cite{noauthor_chipsgov_2022}\cite{solution_video_nodate}. This program promotes the integration of diverse designs and technologies within the domain of microelectronics, ensuring reliability and security. Despite the significant push in bringing the chip fabrication onshore, it's also essential for the same amount of attention to building onshore capabilities of advanced packaging in the US to fully secure the semiconductor supply chain. However, the US currently accounts for only 3\% of the total advanced packaging market share, and the rest of the packaging is done offshore, which may cripple the semiconductor supply chain\cite{lapedus_expanding_2022}\cite{vashistha_toshi_2022}\cite{noauthor_big_nodate}\cite{khan_secure_2022}. Developing advanced packaging capabilities in the US is one of the major and necessary steps to secure the semiconductor supply chain. To this end, each stage of the advanced packaging supply chain, as well as the major actors in each stage, needs to be examined critically to fully understand the current structure of the US-based advanced packaging ecosystem and to identify the bottlenecks that could potentially pose a threat to the stability and security of this ecosystem.

In summary, we analyze the present state of the US-based advanced packaging ecosystem and the possible countermeasures to mitigate the hardware security issues posed by the current advanced packaging supply chain. Our contributions are:
\begin{itemize}
\item We analyze the US on-shore advanced packaging manufacturing capabilities.
\item We investigate the dependency of the US's advanced packaging capabilities on offshore resources in various application sectors of importance.
\item We provide an overview of the current business development initiatives of semiconductor wafer manufacturing in the US and emphasize the necessity for similar initiatives focusing on advanced packaging operations specifically.  
\item We identify the major requirements to develop a secure US-based advanced packaging supply chain.
\end{itemize}

This paper primarily analyzes the US-based advanced packaging supply chain ecosystem and its capabilities (Fig. 2.). The paper's organization is as follows: Section II provides the information on the background and motivation of HI and advanced packaging. Section III introduces the US-based advanced packaging supply chain. Section IV examines some economic and business development aspects of moving advanced packaging operations to the US. Section V provides a roadmap for securing an onshore US-based advanced packaging supply chain. Finally, Section VI concludes the paper.

\section{Background}
\subsection{Motivation of HI}
To keep pace with the continuation of More-than-Moore (MtM) progress, HI is essential to increase performance and yield rate in the smaller nodes, lower power consumption and cost of the semiconductor, and reduce latency among the chiplets. There are several significant drivers for the semiconductor industry to advance HI actively:

\subsubsection{Characteristics of HI}
DARPA identified the three primary drivers to innovate in HI\cite{noauthor_eri_nodate}. Firstly, HI enables \emph{technological diversity} by integrating chiplets from different technology nodes and foundries onto a common interposer. This allows the combination of chiplets with varying technology levels, facilitating the integration of newer and older chiplets into the same package. Secondly, HI supports \emph{functional diversity} by integrating chiplets with different functions onto a single package. This enables the design of SiP solutions that incorporate memory, logic, analog I/O, and MEMS sensor chiplets, allowing for modular and custom designs. Lastly, HI allows for \emph{materials diversity}, meaning chiplets can be made from different materials as long as they do not adversely impact the system's functionality. This flexibility enables optimization of chiplets for specific functions and enhances capabilities using newer materials. Overall, HI promotes the integration of chiplets with diverse technologies, functions, and materials, leading to improvements in performance, cost-effectiveness, and design flexibility in semiconductor systems.
\begin{figure*}[htp!]
    \centering
    \includegraphics[width=17.5cm]{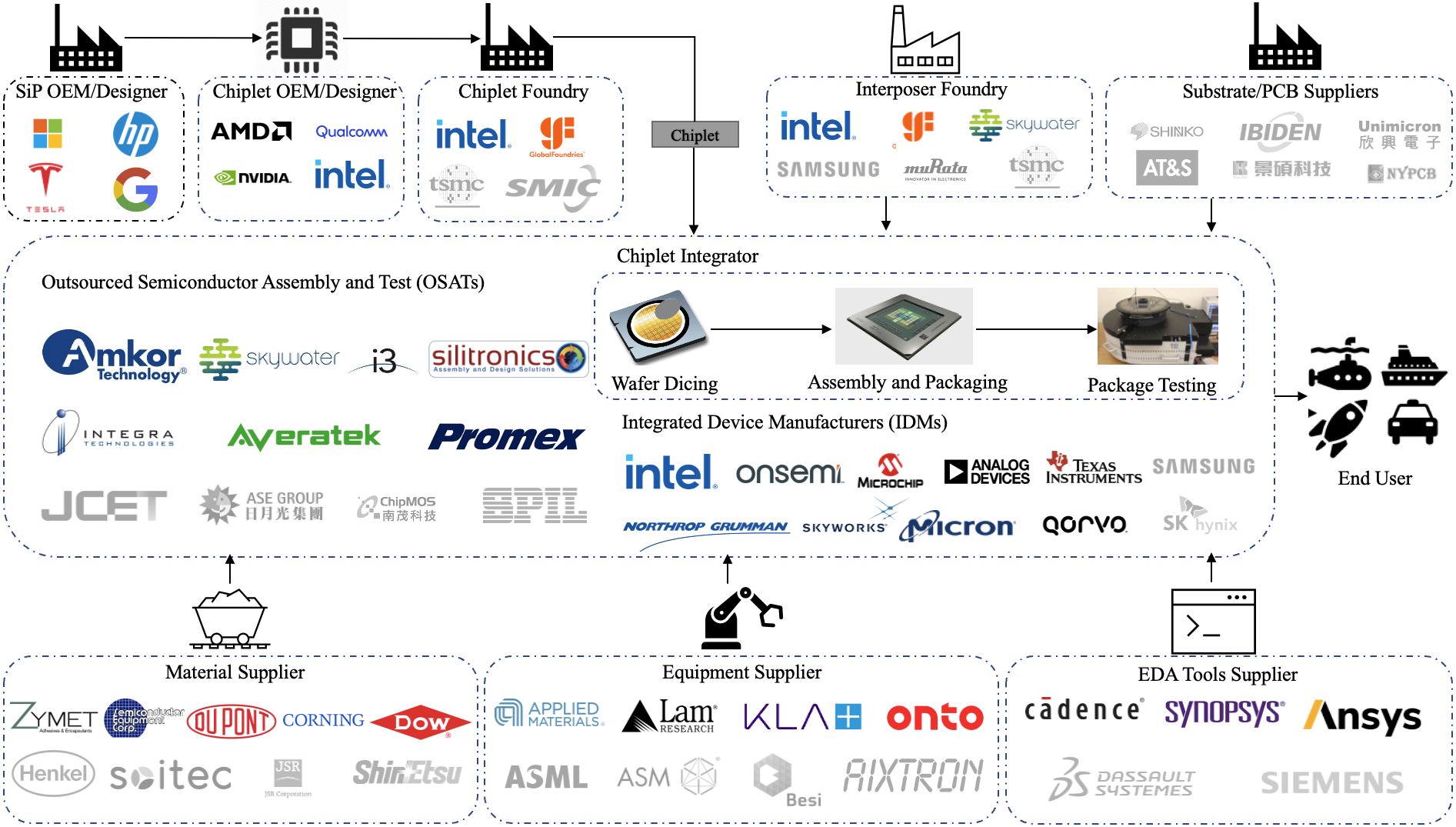}
    \caption{US-based advanced packaging supply chain with onshore and offshore (greyed out logo) companies.}
    \label{fig:galaxy}
\end{figure*}
\subsubsection{Continuance of Moore's Law}
Moore's Law has been the cornerstone of innovation in the semiconductor industry for decades, propelling advancements by doubling transistor density in IC every two years. However, there is growing skepticism regarding the ongoing applicability of this law due to obstacles encountered in scaling transistor sizes (e.g.,  quantum phenomena) and escalating manufacturing expenses\cite{chau_process_2019}. Consequently, novel strategies like HI have emerged, fundamentally transforming packaging and design approaches and presenting a new lens through which to view Moore's Law. Rather than exclusively emphasizing transistor density, these innovative techniques prioritize functional density as a performance indicator, leading to fresh perspectives, valuable insights, and more accurate predictions within the industry.

\subsubsection{Enhancing yield to achieve cost reduction}
By integrating known good dies or chiplets with a higher manufacturing yield, HI promises to enhance yield in SiP solutions. Technological advancements have played a role in improving integration and stacking yields while reducing manufacturing and research costs. To increase electrical die yields and transfer bonding, collective die-to-wafer bonding has been suggested\cite{yang_yield_2011}. Moreover, utilizing chiplets from matured process nodes in SiP development reduces the necessity for post-silicon validation, leading to decreased development costs. Existing research shows promise of superior reliability and high yield in manufacturing high-performance 3D-ICs too\cite{uhrmann_heterogeneous_2018}.

\subsubsection{Minimizing form factor}
Adopting 2.5D and 3D packaging techniques has resulted in smaller form factors and reduced size requirements. This size reduction is achieved by integrating multiple dies into a single package, eliminating the need for separate connections using traces on a Printed Circuit Board (PCB). As a result, the interconnections in these integrated technologies are smaller, leading to improved speed and lower power consumption. 2.5D packaging achieves a higher functional density than legacy packaging, where multiple dies are placed side by side on the top interposer. Still, it falls short of the density achieved by 3D packaging in terms of functional density as it involves stacking dies vertically. However, it also poses challenges in managing thermal issues caused by the heat generated within the stacked dies. Therefore, using 2.5D and 3D packaging techniques in HI enables the realization of a compact form factor, enhanced performance, high manufacturing yields, and reduced overall area requirements of the chip\cite{chau_process_2019}.

\subsubsection{Utilizing SiP technology to enhance performance}
As the performance gains from increasing transistor density on a single-die approach plateaued, HI can continue Moore's Law by incorporating multiple dies in SiP. This integration presents an opportunity for higher performance through improved memory access speeds. For example, 3D packaging technology enables the stacking of CPU and memory dies, leading to enhanced memory bandwidth and reduced transmission latency thanks to shorter interconnects between the dies\cite{li_chiplet_2020}. Ongoing research is focused on improving the communication quality and interconnect in interposers. There are even proposals to introduce active interposers, which embed transistor-based logic circuits within the interposer itself, further enhancing the functional density of the SiP.

\subsection{Advanced packaging to enable HI}
The necessity of integrating multiple dies or chiplets within a single package draws significant attention to the development of advanced packaging technologies, placing semiconductor engineers and physicists at the forefront of their capabilities. Accommodating an increased number of silicon dies within a reduced footprint necessitates both vertical and horizontal stacking, involving additional wire bonding, densely packed smaller bumps, heightened routing complexity, and potential interference from neighboring signal paths\cite{noauthor_heterogeneous_nodate-4}\cite{li_chiplet_2020}. Various packaging technologies have emerged to address these wide design challenges for advancing HI. A few common packaging technologies are discussed below:

\subsubsection{Legacy packaging}
Legacy packaging refers to traditional packaging methods that have been widely used. These techniques include dual in-line package (DIP), quad flat package (QFP), and small outline integrated circuit (SOIC). While legacy packaging has served the industry well, it has certain limitations in size, power dissipation, and signal integrity. However, legacy packaging still finds application in specific areas, such as low-cost devices or applications with lower performance requirements\cite{lau_semiconductor_2021}.

\subsubsection{Flip-Chip packaging}
To enhance the performance and efficiency of the chip and reduce the interconnection, it is necessary to position the chips closer together. Flip-chip Ball Grid Array (FCBGA) based packaging has been developed to achieve this. In this configuration, the chips are placed, or the antenna is formed on the surface of the package. At the same time, the digital, analog, or Radio Frequency ICs are integrated into the bottom of the ball grid array substrate in a monolithic manner. This approach enables improved power efficiency and high data rates. However, it is essential to note that thermal management becomes a challenge when employing this method. To improve connectivity and reduce parasitism, micro-bump is introduced \cite{lau_semiconductor_2021}. 

\subsubsection{Wafer Level Packaging (WLP)}
Wafer-level packaging (WLP) is one of the standard chip packaging approaches where thin metal layers are used to create redistribution layers (RDL). Additionally, fan-out wafer-level packaging (FOWLP) has emerged as a popular method for mmWave microelectronics packaging. FOWLP offers substantial benefits by reducing the size and thickness of the package, resulting in a substrate-less design that enhances RF performance and provides greater design flexibility. However, warpage poses a significant challenge in FOWLP due to the utilization of materials with different coefficients of thermal expansion (CTE). To address this, embedded Wafer Level Ball Grid Array (eWLB) has become a favored FOWLP technology for high-volume production at a reasonable cost\cite{noauthor_heterogeneous_nodate-5}\cite{lau_semiconductor_2021}.

\subsubsection{2.5D Packaging}
In 2.5D packaging, a separate interposer layer is positioned between the chiplets and the packaging substrate. The prevailing trends in packaging applications primarily focus on integrating cutting-edge logic and memory elements within a single package. In this scenario, the interposer's main function is facilitating high-speed data communication between these devices.

\begin{figure}[htp]
    \centering
    \includegraphics[width=8.5cm]{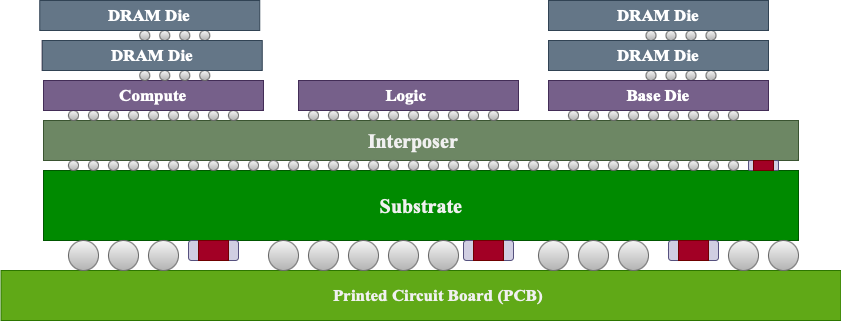}
    \caption{A typical example of 3D packaging.}
    \label{fig:galaxy}
\end{figure}

A great example of 2.5D packaging is Chip-on-Wafer-on-Substrate (CoWoS), developed by TSMC. In CoWoS, multiple chiplets or dies are stacked atop a silicon interposer, which is attached to a substrate using micropumps or through-silicon vias (TSVs)\cite{noauthor_wafer-level_nodate}. Initially, the chips are bonded to the interposer using flip-chip or wire bonding techniques. Following this, the interposer is attached to a substrate. The chip-facing side of the interposer typically comprises multiple metal layers or RDL, which distribute electrical signals from the I/O pads on the chiplets to their corresponding pads on the silicon interposer. Interposer-based 2.5D packaging is particularly valuable when integrating dissimilar chiplets, such as memory, processors, and MEMS-based sensors. Various companies have developed their own 2.5D packaging solutions, encompassing memory, processors, and MEMS-based sensors. Diverse materials have also been employed to create interposers, including IBM's direct bonded heterogeneous integration (DBHi), TSMC's local silicon interconnect (LSI), and ASE's stacked Si bridge fan-out chip-on-substrate (sFO CoS), etc.

Another approach to 2.5D packaging involves using "bridges" to establish connections between adjacent chips. EMIB (Embedded Multi-die Interconnect Bridge) is an example of bridge-based 2.5D packaging \cite{noauthor_embedded_nodate}. In this method, the bridge is fabricated separately and embedded within the cavity of the packaging substrate. Some literature categorizes this bridge-based advanced packaging solution as a 2.3D packaging structure. Silicon blocks with high I/O density are employed to create these "bridges," facilitating interconnections between chiplets positioned in close proximity to each other. Companies are actively exploring developing their own bridge solutions due to the cost-effectiveness of this approach compared to interposer-based 2.5D packaging.

\subsubsection{3D Packaging}
3D packaging technology facilitates stacking and interconnection of semiconductor dies by placing one on top of another, typically employing vertical connections known as TSV (Fig. 3). This approach is commonly employed to stack memory atop a processor or to integrate analog and digital circuits. Intel's Foveros \cite{noauthor_foveros_nodate} is a prime example of 3D packaging technologies, which is a type of Die-on-Die (DoD) packaging, where distinct functional dies are stacked using TSVs and micro-bumps to establish electrical connections between the layers. Another form of 3D packaging is Package-on-Package (PoP), which typically involves vertically connecting two packaged dies and linking them through package vias (TPV). PoP technology finds extensive application in imaging sensors and chips used in portable devices\cite{noauthor_heterogeneous_2023}.

\begin{figure}[h]
    \centering
    \includegraphics[width=9cm]{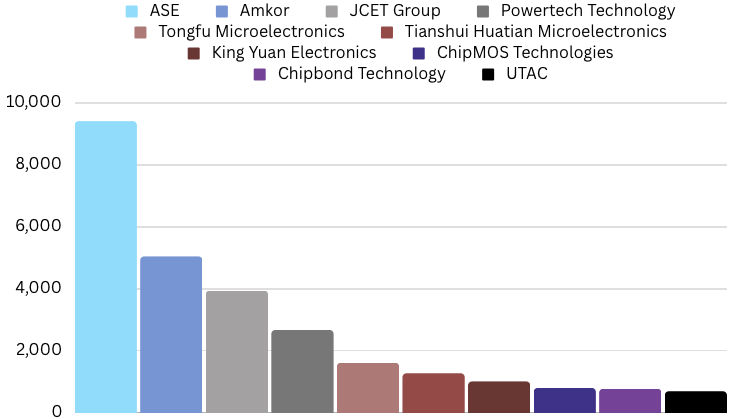}
    \caption{2020 advanced packaging market share in US\$ million.}
    \label{fig:galaxy}
\end{figure}

\section{US-based advanced packaging supply chain}
By introducing the CHIPS (Creating Helpful Incentives to Produce Semiconductors) Act\cite{house_fact_2022}, the US government has signaled its desire and commitment to bring semiconductor wafer fabrication facilities onshore. However, to fully secure the entire semiconductor supply chain, back-end operations cannot be ignored, and it is of utmost importance to pay equal attention to developing advanced packaging capacity onshore as well\cite{rep_ryan_hr4346_2022}. A critical examination of the existing packaging capacity in the US reveals that there is sufficient capacity for legacy packaging. Thus, developing OSATs with advanced packaging capabilities makes more sense. However, with merely a 3\% market share in the global market, the US lags in advanced packaging capacity. Thus, ramping up the advanced packaging manufacturing capabilities in the US is important. To realize the vision of a healthy, burgeoning domestic advanced packaging supply chain, it’s important first to examine the current structure of the advanced packaging ecosystem, capabilities, off-shore dependency, and potential weaknesses in the US. Fig. 2. identifies the agents in the advanced packaging ecosystem in the US and provides an overview of the relationships among these agents. In what follows, the roles of these entities are described, and illustrative examples of the prominent onshore and offshore (greyed out logo) players are provided. 

\subsubsection{SiP OEM/Designer}
Original Equipment Manufacturer (OEM) refers to a company that designs and manufactures products or components used in another company's end product. SiP designers or OEMs either sets specifications for their SiP according to their needs or directly procure the SiP from the chiplet designer. For example, Microsoft buys chips from chiplet designers like AMD, NVIDIA, or Intel for their data centers. On the other hand, Tesla may design a SiP and outsource them to a chiplet OEM for fabrication or procure chips directly from Intel, AMD, and Texas Instruments for their autonomous vehicles. 

\subsubsection{Chiplet OEM/Designer and Chiplet Foundry}
A chiplet designer designs the chiplet for the SiP OEM/designer. The chiplet designer may have a fab or be fabless. In the latter case, a fabless chiplet designer outsources chip fabrication to an onshore or offshore Chiplet Foundry for fabrication. 

AMD, NVIDIA, Apple, Qualcomm, and Broadcom are prime examples of fabless chiplet designers which do not have manufacturing capabilities. For example, AMD, Apple, and NVIDIA outsource chip fabrication to TSMC, an offshore Chiplet Foundry. AMD works closely with TSMC, ASE, and Tongfu Microelectronics for advanced packaging capabilities, such as 2.5D packaging \cite{published_amd_2022}\cite{noauthor_ase_nodate}. Apple entirely depends on TSMC for chip fabrication and advanced packaging capabilities. AMD and NVIDIA datacenter's GPU depend on TSMC for their CoWoS packaging technology. Other prominent Chiplet Foundry are GlobalFoundries, Samsung, SMIC, UMC, etc., which provide chip fabrication services for the fabless chip designer. Due to the current geopolitical tensions between China and Taiwan, the world semiconductor supply chain is severely vulnerable to potential geopolitical turmoil. Such a conflict would affect leading semiconductor companies, such as AMD, NVIDIA, and Qualcomm, among others, as TSMC accounts for 56\% of the total semiconductor market share in the world\cite{published_tsmc_2023}\cite{lapedus_foundry_2021}. Hence, more investment is needed in the US to secure chip production and/or advanced packaging capacity for fabless American chip designers.
\begin{figure*}[htp]
    \centering
    \includegraphics[width=17cm]{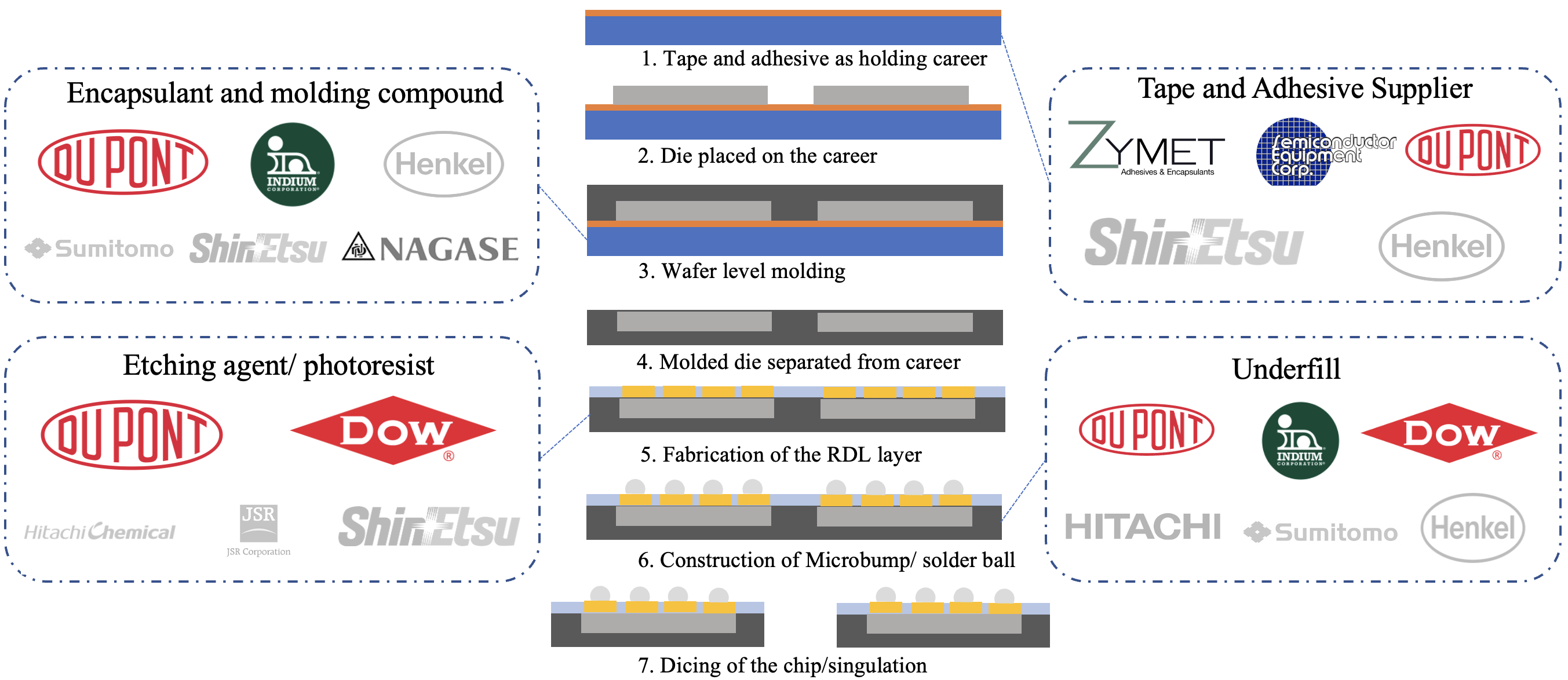}
    \caption{Typical wafer-level packaging process flow with onshore and offshore (greyed out logo) material suppliers}
    \label{fig:galaxy}
\end{figure*}
\subsubsection{Chiplet Integrator} 
A chiplet integrator can either be a single entity offering comprehensive integration services, including interposer design, packaging, assembly, and testing, or it can be divided into multiple separate entities, depending on its expertise or business model. There are primarily two classes of chip integrators: IDMs and OSAT vendors. While an IDM has end-to-end chip design, fabrication, and packaging capabilities, an OSAT vendor only provides packaging design and manufacturing capabilities. 

In the US, a few companies, such as Intel, Qorvo, Texas Instrument, and Onsemi, are qualified as IDMs. For instance, Intel developed the 3D stacking with Foveros technology. Similarly, Qorvo offers advanced packaging capabilities, like an AiP, essential in wireless communication. In addition, Micron is dedicated to memory technologies and related advanced packaging capabilities. Finally, Northrop Grumman and Honeywell are IDMs dedicated to the defense and aerospace sectors. 

There are currently 25 OSAT vendors in the US, but not all can offer advanced packaging capabilities. The most well-known US-based OSAT vendor is Amkor, but Amkor does not have any manufacturing capabilities in the US (Fig. 4.)\cite{support_advanced_2021}. Promex is building on-shore advanced packaging capabilities\cite{noauthor_promex_2022}. Sitronics has advanced packaging capabilities, offering an MCM platform with a buildup substrate, and doesn't use silicon interposer technology. Skywater is building fab and packaging facilities to offer advanced packaging capabilities\cite{rocket55_skywater_2021}. The US has always been a leader in developing cutting-edge semiconductor technologies, including packaging. However, more investment is needed to ramp up on-shore advanced packaging operations' design and manufacturing capabilities.

\subsubsection{Material Supplier} 
Semiconductor advanced packaging requires complex manufacturing and process flow, which consists of various stages like dicing and slicing the wafer, placing it into the mold and wire bonding, stacking the chiplet, or encapsulating the chiplet. In different stages of the process flow, various raw materials are required. Dielectric materials, leadframe, gold wire, encapsulant, and molding compound are some of the most crucial materials for packaging the chip\cite{zou_review_2021}\cite{arifin_epoxy_2018}. Adhesives and tapes are needed to hold the wafer during the dicing step of the wafer. These materials are vital in connecting a manufactured chip to its protective package, with the specific process and materials chosen based on the intended application. For example, one approach involves using bond wire to attach the chip to a lead frame, enabling data transfer between the chip and external devices. A protective ceramic package, plastic substrate, or encapsulant resin can also be bonded to the chip. To attach the chip to packages or substrates, die attach materials like polymers and eutectic alloys are utilized. Etching agents and photoresist chemicals are needed for RDL fabrication\cite{wise_why_2021}. Underfill materials are needed for fabricating the solder ball and microbump\cite{noauthor_heterogeneous_nodate-1}\cite{noauthor_re-shoring_nodate}. 

Fig. 5. explains a typical wafer-level packaging process flow with different materials needed in each step. HD Microsystems (parent company DuPont) provides the chip's etching agent, cleaning solution, and encapsulant for molding and encapsulation. Zymet, Semiconductor Equipment Corporation, and DuPont provide the adhesive and tape needed for holding the chiplet during the dicing step.  Soldering and underfill material for fabricating solder balls and microbumps are provided by Dow Chemical, DuPont, and Indium Corporation. Even though there are suppliers for the necessary raw materials needed for packaging in the US, the majority of the materials are supplied by Japan, China, and Taiwan. The US currently has a 10\% market share for semiconductor material, which may pose vulnerabilities in the semiconductor supply chain, including advanced packaging\cite{noauthor_national_nodate}\cite{noauthor_polymeric_2018}\cite{tang_high_2017}. To support the US-based advanced packaging manufacturing, further investment is required to secure the raw material supply chain.

\subsubsection{Equipment Suppliers}
 Various types of equipment are used in different stages of packaging manufacturing, such as dicing, wire-bonding, microbump, and hybrid bonding. In WLP, the wafer needs to be diced, and then the RDL is formed on top of the wafer. This step requires conventional photolithography equipment that is used for chip fabrication. Flip chip bonding equipment attaches the IC chips directly onto the substrate or PCB by accurately positioning and bonding the solder bumps on the chip to the corresponding pads on the substrate, allowing for high-density interconnections. Wire bonding equipment connects the IC chip to the substrate or PCB using fine wires made of gold, aluminum, or copper. It connects the wires from the chip's bond pads to the appropriate bonding pads on the substrate, enabling electrical connectivity. Die attach machines, place, and bond the IC chips onto the substrate or PCB. They use adhesive materials, such as epoxy or solder, to secure the chip in the desired location. Molding equipment is used in encapsulation, where a protective package is formed around the IC chip. This equipment encloses the chip in a plastic or ceramic material, providing mechanical protection and environmental isolation. Various testing and inspection equipment are used to verify the functionality and quality of the packaged ICs. This includes automated optical inspection (AOI) systems, X-ray inspection machines, electrical testers, and other specialized testing tools. Wafer dicing machines cut the processed silicon wafers into individual IC chips. They utilize mechanical or laser cutting techniques to dice the wafer into the desired chip sizes\cite{tang_high_2017}\cite{noauthor_re-shoring_nodate}.

Applied Materials and Lam Research are prime examples of US-based equipment suppliers for the semiconductor industry and advanced packaging. Besides this, KLA, Onto Innovation, Nordson, Thermofisher Scientific, and Bruker provide various equipment for metrology needs. US companies provide necessary inspection and metrology tools for all stages in the manufacturing process.
\subsubsection{Substrate and PCB suppliers}
The substrate, also known as a package substrate, is a base material on which the IC is mounted. It provides mechanical support, electrical connectivity, and heat dissipation for the IC. The substrate acts as a medium through which electrical signals pass between the IC and the PCB. It is typically made of a high-performance material like ceramic or organic laminate with conductive traces and vias to establish connections between the IC and the external circuitry. The Printed Circuit Board (PCB) is made of a non-conductive material such as fiberglass-reinforced epoxy with a pattern of conductive traces etched onto it. It provides a platform for mounting and interconnecting various electronic components, including ICs. The PCB typically comprises multiple layers of conductive traces and insulating materials sandwiched together. When an IC package is mounted onto a PCB, the package substrate is soldered or attached to the PCB, establishing electrical connections between the IC and the PCB. The PCB provides a means to distribute power and signals to other components on the board, enabling the IC to communicate with other parts of the circuitry.

There is a notable scarcity of IC substrate material within the advanced packaging supply chain. These substrates play a vital role in packaging high-end CPU, GPU, and 5G networking chips manufactured by major companies like Intel, AMD, and Nvidia. Currently, there is no on-shore substrate supplier in the USA\cite{noauthor_five_2022}.

\subsubsection{EDA Tools Suppliers}
Chip design is a highly complex, multi-year process, and without the help of Electronic Design Automation (EDA) tools, it is almost impossible to create chip designs. The importance of EDA tools holds for advanced packaging operations, too. Various types of EDA tools are used to model and analyze the reliability of the packaging, the design of the package antenna, and many other aspects of packaging designs. 

Cadence, Synopsis, Mentor Graphics, and Ansys provided extensive EDA tool options to design the chip and the packaging. For example, Cadence recently started offering their 3D IC SiP simulation tools inventory\cite{noauthor_cadence_nodate}. The same goes for Mentor Graphics, too. With Ansys, we can model and design the Antenna in AiP. Thus, major US EDA tool companies provide all the necessary tools for designing advanced packaging. 

\begin{table*}[ht!]
    \centering 
    \label{packaging requirements}
    \caption{Design requirements in advanced packaging for different application fields.}
    \centering
    \begin{tabular}{|c|>{\centering\arraybackslash}p{13cm}|}
    \hline
            \textbf{Application fields} & \textbf{Advanced packaging requirements} \\
    \hline
            Aerospace and Defense & 
            \begin{itemize}[noitemsep, left=0pt,topsep=0pt, partopsep=0pt, parsep=0pt, itemsep=0pt, after=\vspace{-\baselineskip}]
            \item High reliability, durability, and long product life cycle in the extreme environment
            \item Secure domestic supply chain and leverage verification technologies to safeguard data and system security 
            \item High-performance design for ensuring a competitive edge 
            \end{itemize} \\
     \hline
            5G Communication & 
            \begin{itemize}[noitemsep, left=0pt, topsep=0pt, partopsep=0pt, parsep=0pt, itemsep=0pt, after=\vspace{-\baselineskip}]
             \item Enabling beamforming by embedding antenna in the package for mmWave communication
            \item Low noise, interference, minimizing cross-talk between circuitry
            \item Power efficiency in mobile 5G applications
            \item High-Frequency operation 
            \item Considering temperature issues generated from the high-frequency operation 
            \end{itemize} \\
    \hline
            High-Performance Computing & 
            \begin{itemize}[noitemsep, left=0pt, topsep=0pt, partopsep=0pt, parsep=0pt, itemsep=0pt, after=\vspace{-\baselineskip}]
            \item High-speed interconnects in packaging to enable high-speed data transfer between chiplets
            \item Integrating heat spreaders, heat sinks, and thermal interface materials to ensure heat dissipation
            \item Optimizing power distribution, minimizing power losses, and efficient power power delivery
            \item Increasing testability and yield by stacking multiple chiplets
            \item Vertical stacking of memory and compute die for low latency, high-speed data transmission
            \end{itemize} \\
    \hline
             Automotive & 
             \begin{itemize}[noitemsep, left=0pt, topsep=0pt, partopsep=0pt, parsep=0pt, itemsep=0pt, after=\vspace{-\baselineskip}]
            \item High reliability, longer product lifecycle in an extreme environment such as vibration and extreme temperature
            \item Integrating sensors such as MEMS, LiDAR, and RADAR in a single package
            \item High-performance computing for autonomous vehicles
            \end{itemize}
            \\
    \hline
             Internet of Things (IoT) & 
             \begin{itemize}[noitemsep, left=0pt, topsep=0pt, partopsep=0pt, parsep=0pt, itemsep=0pt, after=\vspace{-\baselineskip}]
            \item Low power consumption
            \item Small form factor
            \item Ensuring low cost per unit
            \item Various sensor integration 
            \item RF performance for 5G applications
            \end{itemize}
            \\
    \hline
             Mobile & 
             \begin{itemize}[noitemsep, left=0pt, topsep=0pt, partopsep=0pt, parsep=0pt, itemsep=0pt, after=\vspace{-\baselineskip}]
            \item Integration of various functionalities, such as processor, memory, sensors, and wireless communication modules
            \item Low power consumption and high-speed data transfer
            \item Efficient thermal management and heat dissipation
            \item Ensuring a small form factor is essential
            \end{itemize}
            \\
    \hline
            Medical and Health &
            \begin{itemize}[noitemsep, left=0pt, topsep=0pt, partopsep=0pt, parsep=0pt, itemsep=0pt, after=\vspace{-\baselineskip}]
            \item Excellent electrical performance to ensure accurate data transmission, signal integrity, and power delivery
            \item Miniaturation and small form factor while maintaining optimal performance
            \item Hermetic sealing to prevent contamination while using in a sterile environment
            \item Ensuring biocompatibility for safe medical applications
            \item Reliability, durability, and long life cycle 
            \item Flexible packaging technology for wearable medical devices
            
            \end{itemize}
            \\
    \hline
    
\end{tabular}
\end{table*}

\subsubsection{Interposer Foundry}
As discussed in the previous section, a silicon interposer is a key component in semiconductor packaging technology and is responsible for increased chip performance and functionality. GlobalFoundries and TSMC are key interposers foundries. Among them, GlobalFoundries is the most well-known interposer foundry in the USA. For example, AMD outsources their chiplet fabrication for their Ryzen processor to TSMC, which requires advanced nodes like 7nm for compute die, and I/O dies from GlobalFoundries, which is in older process nodes. GlobalFoundries fabricates the interposer for the AMD Ryzen processor to integrate all the chiplet in different process nodes. Although GlobalFoundries can fabricate the interposer, there is a significant offshore dependency as TSMC is the primary supplier for interposer die\cite{noauthor_amd_nodate}\cite{jerger_noc_2014}. Hence, while the recent announcement of the planned investment by GlobalFoundries in New York is a welcome development, 
the reliance on offshore resources for interposer dies continues to pose a challenge\cite{millington_globalfoundries_2023}.

\subsection{US-based advanced packaging supply chain needs and capabilities in different fields}
\label{subsec:fields}
The following sections will evaluate US companies' current and future abilities to manufacture and integrate technologies leveraging HI and advanced packaging. These evaluations will be used to identify key areas in US advanced packaging technology markets that cannot be supported by domestic manufacturing facilities alone. 

To complete these evaluations, we will be identifying specific requirements for each technology space, determining if US OSATs for each technology space are capable of manufacturing components leveraging advanced packaging methods, identifying the manufacturing dependencies of US technology markets on foreign OSATs, and providing a security assessment for each of these technology markets (Table I, Table II). In what follows, the US advanced packaging capabilities and dependency in various application fields, such as Defense and Aerospace, 5G and Communications, High-Performance Technologies, Automotive, Internet of things, Mobile, Medical, and Health, are analyzed.

\subsubsection{Defense and aerospace}
The Aerospace and Defence (A-D) sector encounters distinctive challenges attributed to several specific characteristics to enable HI, such as ensuring high-reliability durability in packaging and a secure domestic supply chain (Table I). There are 83 trusted suppliers with onshore manufacturing facilities to secure the semiconductor supply chain for the US defense industry. These suppliers are considered DoD Tier 1 trusted suppliers\cite{noauthor_department_nodate}. It includes fabrication facilities, OSAT, IDM, SiP designer, etc. For example, Qorvo is a semiconductor company dedicated to meeting the consumer market that also designs various application-specific IC (ASIC), RFIC, and communication chips required by the RADAR application for the defense and aerospace sectors. They design, fabricate, manufacture, and package everything in one secure place in the USA to ensure maximum security and reliability for the defense industry needs\cite{noauthor_triquint_nodate}. Defense companies like Raytheon, Honeywell, and Northrop Grumman can also be qualified as IDM with advanced packaging facilities in the USA to meet the needs of the defense industry. So, the USA has significant on-shore capabilities for the A-D sector to meet the demand \cite{noauthor_heterogeneous_nodate-1}.
\subsubsection{5G communications}
The proliferation of $5$G technology through smartphones and Internet of Things (IoT) connected systems has increased data traffic and demand for faster data transmission. High-bandwidth, low-latency applications of 5G networking are responsible for providing real-time feedback from critical systems in the healthcare, transportation, and power industries (Table I). AiP integrates the antenna structure and RFIC dies into a single package to enable beamforming, which is essential for 5G wireless communication.  AiP allows the embedding of a beamforming antenna array into the package itself. AiP allows for a smaller device footprint and shorter wiring distance between the RFIC and Antenna. AiP is particularly essential in enabling high-frequency applications since the antennas are small. For 5G, it is most relevant for mmWave 5G as it utilizes a higher frequency band; therefore, shorter interconnection and tight antenna integration are essential\cite{fischer_77-ghz_2014}\cite{watanabe_review_2021}. 

\begin{figure}[h]
    \centering
    \includegraphics[width=8cm]{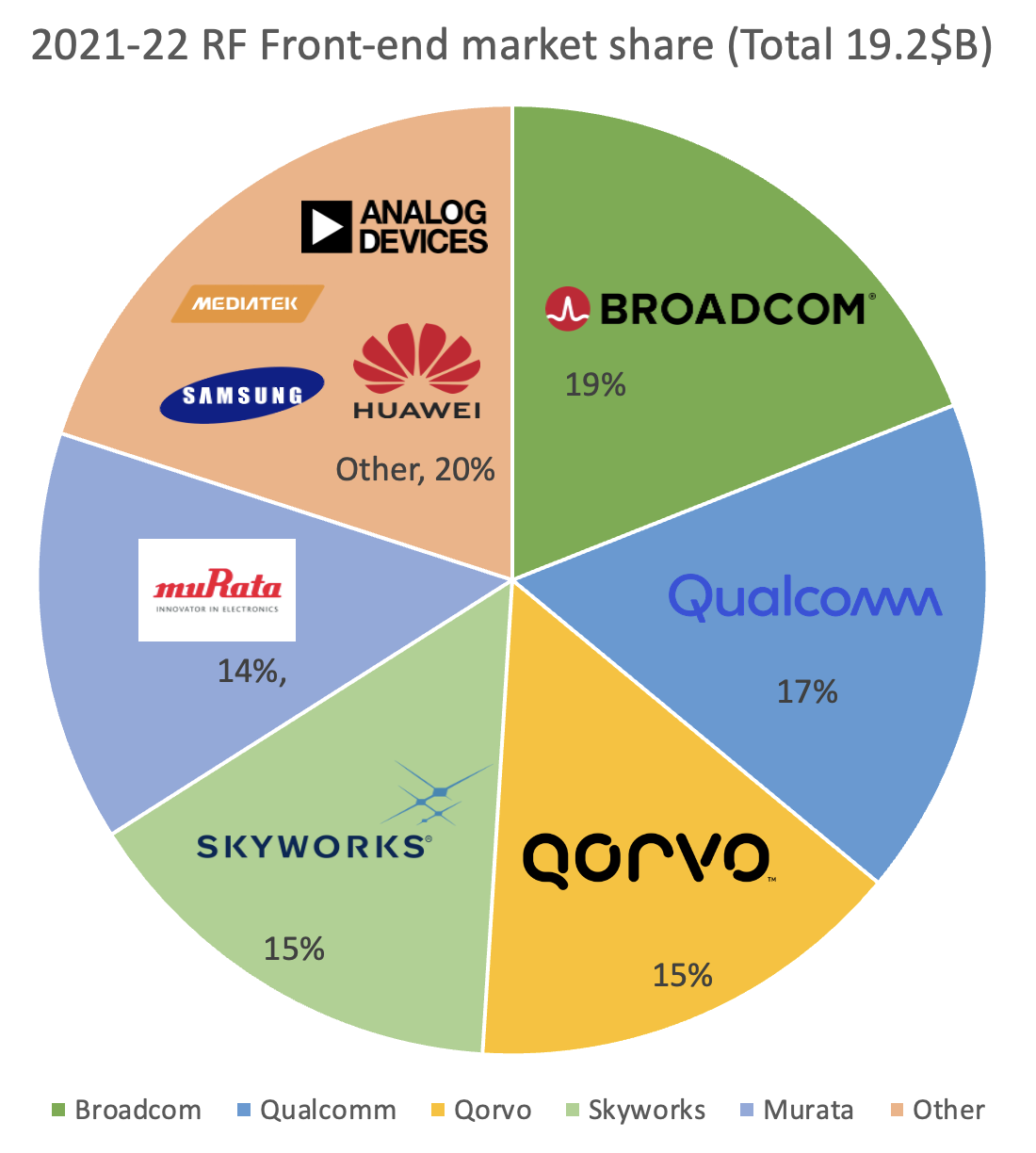}
    \caption{Top 5G chip suppliers in the world.}
    \label{fig:galaxy}
\end{figure}

Few companies in the USA can manufacture the AiP and related RFIC chips to meet the demand for the US 5G market. Qoorvo, Texas Instrument, Analog Devices, Skyworks Solution, and Onsemi are prime examples of developing essential chips and packaging for 5G applications. Only Qoorvo, Skyworks Solution, Analog Devices, Broadcom, and Qualcomm offer AiP essential for enabling beamforming for 5G communication. Among them, only Qoorvo and Skyworks Solution have onshore AiP manufacturing facilities. Infineon and Renesas have onshore foundries, but they are Europe-based companies. The rest of the IDM, like Raytheon, Northrop Grumman, and Honeywell, develop and manufacture AiP and similar solutions for the defense industry for aerospace and military RADAR applications that are not available to meet consumer market demand\cite{noauthor_department_nodate}\cite{molitor_packaging_2020}. Thus, for 5G applications, there is still a sizable offshore dependency to meet the demand for the USA (Fig. 6.)\cite{barrier_telecom_2023}.

\subsubsection{High-performance computing}
The requirements for advanced packaging in high-performance applications are focused on achieving optimal performance, reliability, and efficiency (Table I). In the USA, Intel, AMD, and NVIDIA are the primary suppliers of high-performance chips. Only Intel qualified as IDM can manufacture 2.5D/3D stacked chips onshore. But recently, they lag in semiconductor process node technology compared to off-shore foundries and IDMs like TSMC and Samsung. AMD and NVIDIA are fabless companies that entirely depend on offshore fabs like TSMC and Samsung for their advanced process node up to 5nm. There is an increasing demand for ASIC for accelerating ML workloads like Google's Tensor Processing Unit (TPU) or using FPGA from Altera (acquired by Intel) and Xilinx (Acquired by AMD). These companies depend largely on offshore IDMs like TSMC to fabricate their chips. Besides this, supercomputer and data center is one of the major areas of high-performance computing, which needs a significant amount of processors manufactured in cutting process nodes from TSMC. Network interfacing card (NIC) is essential for supercomputing and data center applications primarily provided by Mellanox Technology, an Israel-based company\cite{schirrmeister_four_2020}. Again, these NICs are fabricated in the TSMC. The ongoing geopolitical tension between China and Taiwan can accelerate possible conflicts in the future that will seriously cripple the semiconductor industry, especially in the high-performance computing application. TSMC and Samsung are building fabs onshore to cope with the demand. The USA still needs to invest more into onshore IDM and OSATS to fully secure the semiconductor supply chain \cite{lapedus_shortages_2017}\cite{noauthor_coming_nodate}\cite{noauthor_heterogeneous_nodate-2}. Besides this, the Ajinomoto Build-up Film (ABF) substrate is crucial for high-performance computing applications. The lack of onshore ABF substrate production may pose a significant supply chain risk for high-performance applications.  

\subsubsection{Automotive}
The requirements for advanced packaging in automotive applications focus on critical factors such as high reliability, longer product life cycles, and sensor integration (Table I). In late 2020, as the automotive market began to recover, car suppliers faced difficulties regaining their production capacity. Re-establishing this capacity has proven to be a challenging task. Most automotive ICs utilize wire-bond packages, including SOIC, TSSOP, QFN, QFP, BGA, and Power Discrete. According to Yole, 90\% of packaging for automotive applications consists of lead or laminate substrates. Substrate supply shortage was one reason for the automotive chip shortage \cite{lapedus_shortages_2017}. Besides this 2.5D, 3D packaging is also needed for high-performance processors in autonomous vehicles for AI workloads. High-performance processors have an offshore dependency on TSMC and Samsung. AiP is needed for mmWave RADAR, which is one of the primary sensing modalities for autonomous vehicles. Offshore players like Valeo, Robosense, and Livox dominate most of the LiDAR supply. Velodyne is a US-based LiDAR supplier that only accounts for 3\% of the LiDAR market share. However, for LiDAR and RADAR packaging, there is a significant offshore dependency. Amkor is the only US-based OSAT to provide packaging solutions for automotive sectors, but they don't have any onshore manufacturing capability. Thus, a significant offshore dependency for packaging needs to be addressed to secure the US-based automotive chip ecosystem\cite{noauthor_lidar_2021}\cite{support_advanced_2021}\cite{noauthor_heterogeneous_2023-2}\cite{noauthor_components_nodate}\cite{noauthor_automotive_2022}.

\subsubsection{Internet of things (IoT)}
With more than 500 billion devices expected to be connected to the Internet by 2030, the IoT requires more heterogeneous systems to advance these applications further. Ensuring low cost per unit, power efficiency, small form factor, and sensor integration are some crucial requirements while designing packaging solutions for IoT (Table I). This will allow many common-use devices to connect the world further, increasing ease of life daily. IoT devices are used in various applications such as medical and health, wearables devices, edge AI, autonomous vehicles, etc. AiP is needed for 5G connectivity in IoT applications. The fan-out package is widely used for edge AI applications in IoT devices to enable high-density connection in the packaging. IoT devices commonly use FOWLP for small form factor packaging. There are significant uses of IoT in medical applications, such as detecting metabolites to diagnose health problems quickly. Besides, various chips for different IoT applications largely depend on legacy packaging. Thus, diverse packaging technologies for IoT application has a sizable offshore dependency, which needs to be addressed \cite{noauthor_heterogeneous_nodate-1}. 

\begin{table*}[ht!]
\caption{US-based advanced packaging supply chain dependency and issues.}
\label{Table comparison}
\centering
\begin{tabular}{|c|c|c|c|c|}

    \hline
            \textbf{\thead{Application field}} & \textbf{\thead{Type of packaging\\ technology}} & \textbf{\thead{Offshore\\ dependency}}& \textbf{\thead{US-based \\IDM/OSATs}} & \textbf{\thead{Dependencies\\ and issues}}\\
    \hline
            \thead{Aerospace and Defense} & \thead{AiP, RADAR packaging}  & Low  &\thead{Qoorvo,\\Honeywell,\\Northrop Gruman} &\thead{Secured by 83 trusted onshore suppliers,\\high manufacturing cost, low volume} \\
    \hline
            \thead{5G Communication}  &  \thead{Legacy packaging, \\AiP} & Medium & \thead{Qoorvo,\\ Analog Devices,\\Skyworks} & \thead{Substrate dependency}  \\
    \hline
            \thead{High-Performance\\ Computing} & 2.5D, 3D & High & \thead{Intel}&\thead{Highly dependent on offshore IDM and OSATs\\ like TSMC for advanced node} \\
    \hline
             \thead{Automotive} & \thead{Legacy packaging, 2.5D/3D,\\ WLP, FCBGA, AiP} & High & \thead{Texas Instrument} &\thead{Substrate dependency, significant offshore \\ dependency for diverse packaging technology \\for RADAR, LiDAR, sensor integration }  \\
    \hline
             \thead{Internet of Things (IoT)} & FOWLP & High & \thead{Texas Instrument,\\Marvell}& \thead{High offshore dependency}\\
    \hline
             \thead{Mobile} & PoP, WLP, Stacked dies &  High&\thead{Skyworks,\\ Qoorvo} & \thead{Highly dependent on offshore IDM and OSATs\\ like TSMC for advanced node} \\
    \hline
            \thead{Medical and Health} & WLP, FCBGA & High & \thead{Analog Devices} & Substrate dependency \\
    \hline
\end{tabular}
\end{table*}
\subsubsection{Mobile}
The mobile sector is one of the major catalysts for innovation in electronics. It represents a large portion of today's electronics market and, most importantly, is used by 75\%-80\% of the world's population. Sensor integration, low power, small form factor, and efficient thermal management are primary considerations while designing packaging solutions for mobile applications (Table I). To continue meeting the demands of mobile devices becoming smaller and faster, the packaging within these devices must also continue to improve. The SoC used in mobile could be PoP architecture or WLP, and the memory chip could be stacked architecture. PoP solutions have found widespread application in baseband and applications processors found in mobile phones. High-end smartphones, in particular, have rapidly adopted PoP packaging due to their ability to meet their demanding requirements for I/O and performance. One of the key advantages of stacked PoP is that individual devices can undergo thorough testing before assembly, ensuring optimal functionality. Many sensors and RF chips used in mobile also use SiP technology. Qualcomm and Apple are the most prominent mobile SoC providers in the USA. However, they entirely depend on TSMC's advanced node to fabricate their SoC and packaging solution. Skyworks Solution or Broadcom could provide sensors or communication chips. Skyworks is an IDM, which is a US-based company. As the mobile SoC entirely depends on TSMC or Samsung, it poses a significant security threat in the supply chain\cite{pushkar_aptew_advanced_nodate}\cite{hsieh_advanced_2016}\cite{noauthor_heterogeneous_nodate}.

\subsubsection{Medical and Health}
The medical and health sectors are critical application fields for semiconductor chips. Various implants such as a pacemaker, prosthetics, sensors, medical instruments, and machinery are used in medical applications. Miniaturization, biodegradability, and excellent electrical performance are key requirements for packaging solutions for medical and health applications (Table I). Various substrate materials are employed in medical electronics, including ceramic substrates like LTCC, flexible circuits, and laminate substrates like BT resin and FR-4. Rigid substrates, often incorporating glass-reinforced materials, are commonly used alongside flexible substrates\cite{shahane_nanopackaging_2018}. Liquid crystal polymer films are also employed in specific devices. Moving forward, there will be a growing focus on utilizing biocompatible substrates. Regarding cover materials, a solder mask is frequently used for short-term applications, which are Pb-free and made with gold. Additionally, Parylene coatings find utility in various cases. As the US semiconductor packaging ecosystem has a critical offshore dependency for the substrate, this may cripple chip supply for the medical and health sector\cite{noauthor_heterogeneous_nodate}\cite{areej_abf_2020}\cite{noauthor_abf_nodate}. 

\subsection{Emerging packaging and manufacturing technology}
\subsubsection{Additive manufacturing for advanced packaging}
Additive manufacturing, also known as 3D printing, is a manufacturing process that involves building three-dimensional objects by adding material layer by layer. Unlike traditional manufacturing methods that involve subtractive processes such as cutting or machining, additive manufacturing starts with a digital 3D model and uses specialized machines to create the physical object. However, recently, there has been an increasing trend to use additive manufacturing in semiconductor advanced packaging, which offers some lucrative advantages such as flexibility, miniaturization, cost-effectiveness, rapid prototyping materials diversity, etc. The ability to print intricate features and structures at a micron scale enables the development of highly integrated packages, accommodating the trend toward miniaturization in the semiconductor industry. Manufacturers can quickly test newly developed packaging designs and technology with rapid prototyping. 

Averatek and Optomec are prime examples of US-based companies that offer additive manufacturing for advanced packaging. Averatek's A-SAP technology allows for the direct printing of circuits on various flexible substrates, including plastics, polymers, and flexible glass, which eliminates the need for traditional etching processes and enables the production of complex circuitry with fine-line widths and tight geometries\cite{noauthor_-sap_nodate}. Optomec is a leading provider of additive manufacturing solutions that offer WLP, high-frequency RF interconnect in the packaging, and shielding with additive manufacturing techniques. Considering the advantages of additive manufacturing in semiconductor packaging, the US must ramp up this sector to meet future demand\cite{germann_aerosol_2014}\cite{erickson_wafer_2021}. 

\subsubsection{Silicon photonics}
Silicon photonics is a cutting-edge technology that has gained significant attention recently due to its potential for revolutionizing data communication systems. It involves using silicon-based materials to manipulate light, allowing for integrating photonic and electronic components on a single chip. This integration of photonics and electronics offers numerous advantages, particularly in HI.

One key advantage of silicon photonics in HI is its ability to provide fast data rates and wide bandwidth capabilities. Traditional IC packaging faces challenges when connecting many devices to the internet to each other. However, silicon photonics offers the potential for extremely fast data rates and wide bandwidth, enabling efficient and high-speed communication between devices. This is achieved through 3-D stacked packaging, where interconnects are significantly shorter than traditional packaging, reducing latency delays. Furthermore, integrating photonics with other components on a single chip allows for efficient use of space\cite{noauthor_heterogeneous_2023-1}\cite{hargrove_review_2018}. AIM Photonics and Atomica are examples of US-based companies that excel in silicon photonics and are dedicated to related packaging technologies that empower optical communications, 3D sensing, LiDAR, augmented/virtual reality (AR/VR), thermal imaging, illumination, and other types of optical sensing\cite{noauthor_capabilities_nodate}.

\section{Economics of advanced packaging: nationally vs. globally}
\label{sec:economic}
The CHIPS and Science Act of 2022 is aimed to strengthen the American semiconductor industry and ensure US leadership in the design and development of technology solutions in defense and aerospace, 5G and communications, automotive, mobile, medical, and health applications, as well as high-performance computing and IoT systems as discussed in Section \ref{subsec:fields}.
The CHIPS and Science Act appropriated a total of \$52.7 billion in funds to the Department of Commerce, the Department of Defense, the Department of State, and the National Science Foundation to advance the American semiconductor industry by supporting research, development, manufacturing, and workforce development activities. While \$39 billion is allocated to manufacturing incentives, \$13.2 billion is committed to research and development as well as workforce development. Furthermore, the bill established the International Technology and Security Innovation (ITSI) Fund to provide \$0.5 billion for the promotion of secure and trustworthy information and communication technologies and to establish secure and resilient global semiconductor manufacturing supply chains worldwide\cite{noauthor_funding_nodate}. 

To enable the realization of the vision set forth by The CHIPS and Science Act, the CHIPS Program Office announced plans to invest in (i) leading-edge logic devices, (ii) advanced packaging, (iii) leading-edge memory devices, and (iv) current-generation and mature-node semiconductors\cite{noauthor_vision_nodate} 
The CHIPS Program Office recognizes the industry's distinction between conventional and advanced packaging and has identified different strategies for conventional and advanced packaging operations. The difficulties associated with reshoring conventional packaging operations in an economically competitive manner are well recognized \cite{noauthor_vision_nodate}. 
Hence, to ensure adequate and secure global conventional packaging capacity for the US, the emphasis will be placed on shifting conventional packaging activities from countries of concern to US allies and partners. On the other hand, to develop a healthy and robust advanced packaging capacity in the US, multiple high-volume advanced packaging facilities will be established domestically to ensure that the US becomes a global technology leader in commercial-scale advanced packaging for logic and memory devices. 

The reconfiguration of global conventional packaging operations and the development of domestic advanced packaging operations, as well as the potential impact of these two major changes on buyer and supplier relationships in the underlying supply chains that support these operations, can be contextualized using the concept of value chains. 
A \emph{value chain} is comprised of a series of operations required to design, manufacture, and deliver a finished product to a customer.
The operations in a value chain are performed by a set of firms, and it is possible to identify a \emph{lead firm} that drives the chain in terms of value addition and distribution and a set of \emph{suppliers} firms that provide the know-how and capabilities needed by the lead firm to deliver the product to its customers. There exists a stream of business and economics literature that investigates buyer and supplier interactions in value chains. It is possible to distinguish three types of value chains, which are namely local, regional, and global value chains.
\emph{Local value chains} (LVCs) connect lead firms and suppliers within a single country. Similarly, \emph{regional value chains} (RVCs) connect lead firms and suppliers within a single world region that may be defined by common regulatory regimes (e.g., the European Union) or offer preferential trading rules for regional members (e.g., NAFTA or ASEAN), or have a national regional identity (e.g., Latin America). Finally, \emph{global value chains} (GVCs) connect lead firms and globally dispersed supplier firms all around the world.

Today, most complex products are built using GVCs, including semiconductor products. The practice of dividing a value chain into different stages and seeking the optimal location and supply mode for each of these different stages has moved substantial portions of value chains for many products from developed economies (e.g., the US and Europe) to emerging economies (e.g., countries in Southeast Asia)\cite{noauthor_interconnected_nodate}. 
As a result, most of the manufacturing and packaging operations in the semiconductor industry are in countries outside of the US today (see Table III). Furthermore, some production processes are segmented so finely that some products may cross international borders numerous times. For instance, a single chip may require more than 1,000 processing steps and may have to pass through international borders 70 or more times before reaching an end customer \cite{noauthor_globality_nodate}. 
\begin{table}[ht]
\caption{US share in the global semiconductor industry.}
\label{Table semi}
\centering
\begin{tabular}{|c|c|c|}
\hline
\textbf{Stages} & \textbf{US Share} &  \textbf{Global Share} \\ \hline
Design & 85\% & 15\% \\ \hline
Manufacturing & 12\% & 82\% \\ \hline
Packaging & 3\% & 97\% \\ \hline
\end{tabular}
\end{table}

After decades of persistent and pervasive growth and expansion of GVCs, challenges associated with GVCs have been recognized within the past two decades. These challenges relate to the underestimation of total costs associated with global operations, capital costs of inventories in transport,  intellectual property theft, environmental impact, and the need to ensure secure supply chains for critical products, among others. These challenges necessitated the reconfiguration of GVCs.  Reshoring, i.e., bringing operations back to the home country, is one of the practices that lead firms may use to as they redesign their value chains \cite{elia_post-pandemic_2021}\cite{canello_reshoring_2022}. 

Starting with the 2008-2009 global financial crisis, firm-level initiatives for reshoring have proliferated. Clothing and footwear, electronics, automotive, mechanical machinery, and equipment, as well as furniture and home furnishing, are the sectors where such firm-level initiatives were observed\cite{noauthor_unctad_nodate} 
More recently, government-level policy interventions to encourage reshoring have also been observed. The US, the UK, Japan, France, and Italy have been among the national governments that enacted policy tools to encourage reshoring \cite{canello_reshoring_2022}. However, there is also a realization that the vision, leadership, and commitment of national governments need to be supported by the active engagement of local governments to develop locally tailored policy tools to help develop shorter supply chains that rely more heavily on local suppliers with the expertise, capability, technological readiness, and capacity needed by the lead firm \cite{pegoraro_regional_2022}.

Reshoring is not a new concept, and there is a nascent stream of research that examines the economics and business strategy of reshoring \cite{barbieri_what_2018}\cite{fratocchi_motivations_2016}\cite{pedroletti_reshoring_2023}\cite{wiesmann_drivers_2017}. 
It is well known that when reshoring is connected with a drive to enhance operations and upgrade, firms tend to seek local business and research partners that have advanced competencies \cite{pal_competitive_2018} 
in addition to skilled workforce \cite{pegoraro_regional_2022}. A robust and burgeoning network of suppliers and supporting operations may facilitate accelerated innovation and continuous growth \cite{pegoraro_regional_2022}. 
Furthermore, interactions with local universities may not only push the frontiers in material, product, process, and equipment innovation but ensure access to a skilled workforce as well \cite{gadde_future_2019}\cite{gylling_making_2015}.

\begin{figure*}[htp!]
    \centering
\includegraphics[width=18cm]{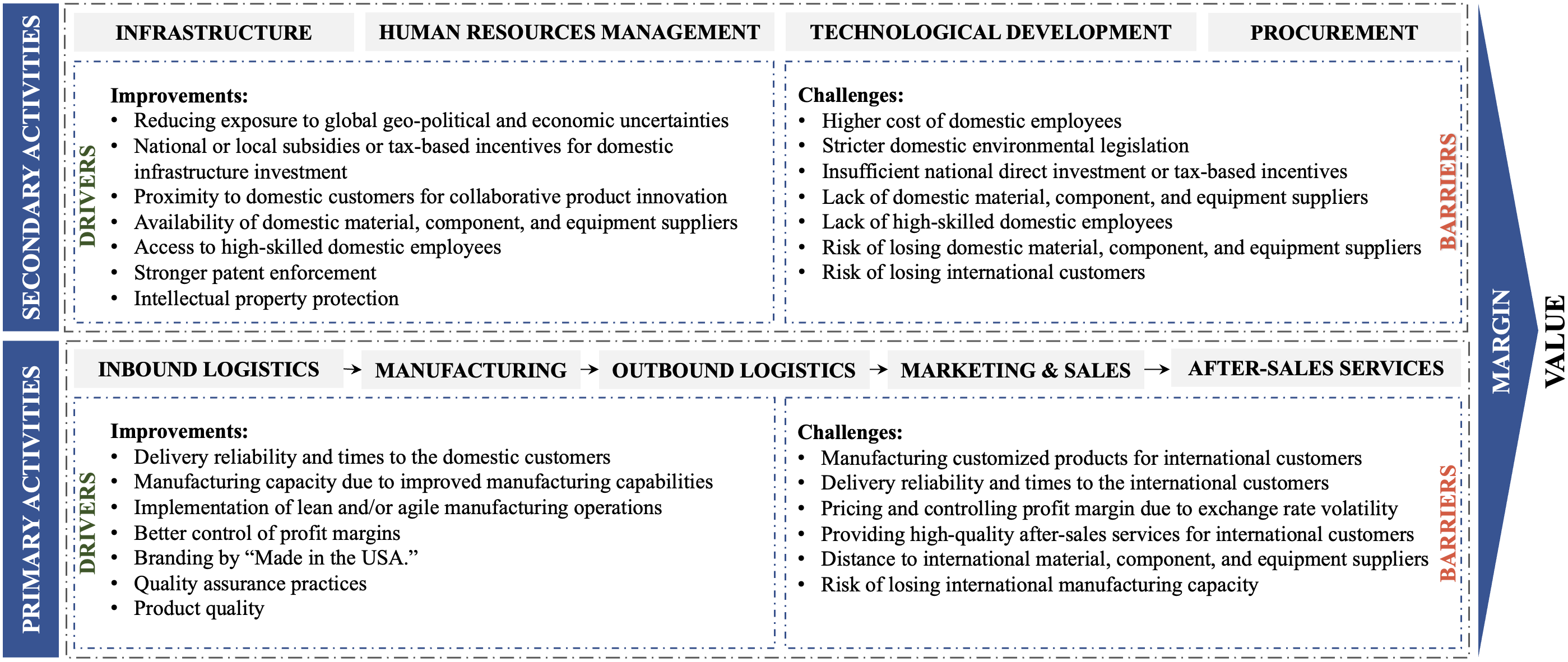}
    \caption{Drivers and barriers to reshoring through a value chain lens focusing on the primary and secondary activities of a firm.}
    \label{fig:galaxy}
\end{figure*}
The reasons why firms reshore have drawn ample attention from practitioners and academicians. While a number of studies investigated country-specific data to examine the reasons for reshoring (e.g., in the US\cite{ellram_offshoring_2013}\cite{noauthor_reshoring_nodate}
and Germany\cite{kinkel_future_2014}), other studies focused on developing theory-grounded interpretations of these motivations \cite{fratocchi_motivations_2016}\cite{srai_institutional_2016}\cite{stentoft_manufacturing_2016}\cite{barbieri_what_2018}. 
Based on a content analysis of peer-reviewed academic journal papers written in English, the drivers and barriers associated with reshoring were broadly categorized into five major groups relating to global competitive dynamics, host country, home country, supply chain, and the lead firm\cite{wiesmann_drivers_2017}. 
The drivers and barriers to reshoring are examined from a value chain analysis perspective as well. Note that all of the activities that make up a firm's value chain can be split into two categories that contribute to the firm's profit margin: primary activities and secondary activities \cite{noauthor_competitive_nodate}.
The primary activities of a firm relate directly to the creation of the product. These are inbound logistics, manufacturing (or operations), outbound logistics, marketing, and sales, as well as after-sales services. Secondary activities of the firm help primary activities to create a competitive advantage for the firm, which can broadly be categorized into infrastructure, human resources management, technological development, and procurement. Fig. 7. provides an overview of drivers and barriers to reshoring through a value chain lens focusing on the primary and secondary activities of a firm. It can be argued these drivers and barriers have to be taken into account by any firm-level initiative for reshoring advanced packaging operations.

In what follows next, a brief snapshot of the current initiatives is provided, and various types of partnerships are highlighted. In addition, the needs that pertain to the development of a national advanced packaging industry are discussed.

\subsection{Current initiatives}
The aftermath of supply chain failures experienced during the COVID-19 pandemic brought some industries to a halt or stymied their growth, and the CHIPS Act has spurred a lot of activity to reshore semiconductor manufacturing operations. Consequently, the construction of semiconductor design and manufacturing facilities is thriving in the US. 
According to Industrial Info Resources \cite{noauthor_us_nodate},
there is currently \$300 billion worth of active semiconductor projects at various stages of development in the US, and New York (\$110 billion), Arizona (\$86 billion), and Texas (\$66 billion) are the three states that are leading the way. While Micron Technology, Inc.\ and GlobalFoundries are the firms investing in New York. Intel and TMSC are investing in Arizona. Similarly, Samsung Group and Texas Instruments are investing in Texas. These investments are expected to boost chip production in the US. 

In the current landscape, most ongoing and planned investments are consolidating around particular geographical clusters in the US. Arizona and Texas, the two historical powerhouses of semiconductor wafer fabrication in the country, are drawing in a lot of expansion and/or development projects for two main reasons. First, these states already have existing well-established semiconductor wafer fab ecosystems. Second, their local governments have demonstrated their support for semiconductor manufacturing by providing incentives and facilitating the process. More recently, Ohio has emerged as an attractive state, with over \$20 billion in investment announced for construction fabs around the Colombus area. Similarly, New York is also offering considerable incentives to encourage fab construction. Among other states that attracted investment are Indiana, Idaho, New Mexico, Oregon, Utah, and Virginia\cite{noauthor_roadmap_nodate}. 

A careful examination of the press releases reveals three types of investment and business development projects. In the first group are projects that are initiated and steered by a single firm (e.g., Micron Technology's DRAM fab at Clay, New York) \cite{noauthor_micron_nodate}.
Another group of projects is independent technology park development and/or expansion projects where a group of firms is creating symbiotic clusters of businesses (e.g., NHanced Semiconductor, Everspin Technologies, Trusted Semiconductor Solutions, and Reliable MicroSystems at 
WestGate@Crane Technology Park in Odon, Indiana) \cite{brown_ground_nodate}.
Finally, a group of projects is the development of new or expansion of technology parks that are co-located with a research university (e.g., SkyWater Technology Foundry at the Discovery Park District at Purdue University at West Lafayette Indiana) \cite{service_skywater_nodate}.

The states that are attracting investments are creating entities to facilitate the process of developing semiconductor ecosystems. For instance, in Indiana, there are two such entities: Indiana Economic Development Corp (IEDC) and Indiana Regional Economic Acceleration and Development (READI)\cite{noauthor_indiana_nodate}. 
To support the WestGate@Crane Technology Park, IEDC is providing support in the form of incentive-based tax credits and training grants, whereas Indiana READI is offering support for infrastructure development. Similarly, to support the SkyWater-Purdue partnership, IEDC is offering conditional tax credits, training grants, redevelopment tax credits, conditional structured performance payments, innovation vouchers, and manufacturing readiness grants. It should be noted that most, if not all, of these incentives are performance-based, i.e., the firms will be eligible to claim state benefits after making eligible investments in activities that foster innovation and after hiring and training employees. 

\subsection{Incentives needs for advanced packaging}
In this context, what appears not to have gained sufficient momentum yet are plans for active investment projects primarily geared toward advanced packaging operations. If a solid and robust advanced packaging ecosystem is not developed in the US, the chips produced by the new manufacturing facilities around the nation will have to be sent to offshore facilities for packaging as before. Then, the CHIPS Act will not have attained its goal of securing semiconductor operations that span the entire supply chain from design to manufacturing to packaging to ensure US leadership in next-generation advanced technology solutions. For instance, TSMC recently announced plans to invest \$2.9 billion into an advanced chip packaging plant in Taiwan to keep up with the increasing demand in the artificial intelligence market \cite{chiang_tsmc_2023}. 
There is a need to secure similar, if not more ambitious, investments focusing on advanced packaging operations in the US.  

First and foremost, it is important to distinguish between two types of advanced packaging operations: rapid prototyping and commercial manufacturing. While advanced packaging capacity for rapid prototyping would ensure US leadership at the forefront of innovation of novel and next-generation advanced packaging products and technologies, advanced packaging capacity for commercial manufacturing would ensure the US to be a major player in satisfying the consumer demand for products that require advanced packaging technologies. Hence, supply chains to support both rapid prototyping and commercial advanced packing operations are needed.
 
As noted earlier, one of three agents in the broader semiconductor ecosystem may perform advanced packaging operations: the IDMs, foundries, or OSAT firms. From a supply chain perspective, while integrating advanced packaging operations into IDMs and foundries is a form of insourcing, using OSAT firms as third-party vendors is a form of outsourcing. The supply chains needed to support these two sourcing operations would require different supply chain configurations. For the former, the IDM or the foundry would have to expand its network of suppliers to provide all the materials, tools, and equipment needed to support advanced packaging operations as well. For the latter, the OSAT vendors would need to cultivate or have access to such networks.

Another factor to note is the need to create infrastructures to enable material, product, process, and equipment technology start-ups to innovate in the advanced packaging space. There is also a need to create infrastructures to enable existing micro and small enterprises to participate in developing supplier networks needed for advanced packaging operations. The role of micro and small enterprises in reshoring has mostly ignored, but there exists convincing evidence on the potential role of micro and small enterprises to support the reshoring of GVCs \cite{canello_reshoring_2022}.
In this regard, developing technology parks that provide clean room manufacturing and packaging capabilities and guaranteed contracting are among the mechanisms that would be particularly important for developing a robust advanced packaging supply chain in the US.

There is insufficient public information on whether any of the wafer fabs or foundries under construction are planning to integrate advanced packaging operations as well. Hence, there is a need to ensure not only the IDMs and foundries plan to develop advanced packaging operations but third-party OSAT firms are encouraged to develop the capabilities, capacity, and readiness in the US to meet the future demand for advanced packaging operations.

\section{Future roadmap for securing onshore US-based advanced packaging supply chain}
\subsection{Supply Chain Issues}
There are many supply chain-related drivers to reshoring (see Section IV), a guarantee to secure a supply chain. However, there is a need to critically examine each stage in the supply chain, from raw materials all the way to the finished good stage, to fully secure a supply chain. To this end, there is a need to fully map a supply chain, including not only the first-tier suppliers but second-tier suppliers and raw material suppliers as well, and identify all stages where the supplier base is limited. It can be argued that the supplier base at a stage of a supply chain is limited if there is a sole or a single supplier at this stage. A supplier is a sole supplier if there are no other suppliers that have the same expertise, capability, technical readiness, and capacity. Similarly, a supplier is a single supplier if other suppliers have the same expertise, capability, technical readiness, and capacity, but the supply chain is configured to make acquisitions from a single supplier. In both cases, the entire supply chain becomes vulnerable to shortages experienced by a sole or a single supplier. Hence, it is critically important for supplier bases at all stages as well as the supply chains of the suppliers to fully characterize the potential vulnerabilities \cite{10238612} of a given supply chain. This is a crucial first step to developing a secure US-based advanced packaging supply chain. In what follows next, some of the most critical bottlenecks that a US-based advanced packaging supply chain may experience.
\begin{figure}[t]
    \centering
    \includegraphics[width=9cm]{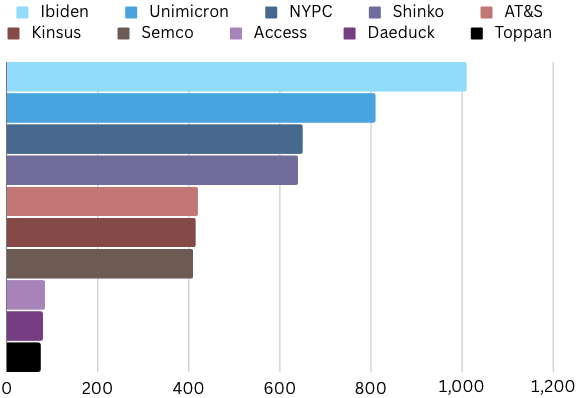}
    \caption{2021 ABF substrates market share in US\$ million.}
    \label{fig:galaxy}
\end{figure}

In this context, a major bottleneck for advanced packaging operations is the total off-shore dependency on IC substrate material, specifically concerning ABF substrates. All the leading ABF substrate producers, such as Ibiden, Unimicron, AT\&S, etc., are offshore companies (Fig. 8.)\cite{noauthor_advanced_2022}. These substrates are crucial for packaging high-end CPU, GPU, Automobile, and 5G networking chips manufactured by major companies like Intel, AMD, and Nvidia. The scarcity has been further intensified by fires at Taiwanese substrate producers in October 2020 and February 2021, aggravating the supply shortage even more. As a result, there have been considerable delays of up to 40 weeks for specific substrates, which is one of the main reasons for the automotive chip shortage. 
Due to the lack of substrate supply, the supply of chips get hampered; thus, the automotive sector also suffers, and vehicle price soar. Lead frame and molding compounds are also one of the major bottlenecks in the whole packaging supply chain, which outside players dominate\cite{noauthor_heterogeneous_nodate-3}\cite{noauthor_intel_nodate}. Several major chipmaking companies, including Intel, AMD, and Nvidia, have provided financial support for approximately 50\% of the capital-expansion endeavors of four prominent high-end ABF substrate manufacturers, namely Ibiden, Shinko, Unimicron, and AT\&S to boost substrate production to address potential substrate scarcity issues \cite{noauthor_abf_nodate}. 

Another major bottleneck is the supply of raw materials like essential chemicals needed for various stages of the packaging process flow, such as wet chemicals, solvents, photoresists, gases, and wafers/substrates. For example, photoresist chemical etching agent supplies mostly depend on offshore. According to the report from the Arizona Commerce Authority (ACA), the US only has 10\% market share for semiconductor materials\cite{noauthor_national_nodate}. The US semiconductor market relies on 31\% of its ultra-high purity chemicals, such as IPA and H2SO4, from Asia, making the industry heavily dependent on long and sometimes fragile supply chains. US semiconductor device production has declined significantly in the past two decades, but with a potential 30\% increase in chip production over the next 3-5 years due to the CHIPS act, materials such as wet chemicals, solvents, photoresists, gases, and wafers/substrates, especially the wet chemicals availability will tighten and could become critical and strain the supply chains in near future unless additional capacity is established. Failure to expand the US supply chain and ramp up onshore production of critical materials may hinder the chip expansion plans and the advanced packaging capabilities \cite{davis_shortage_2021}. 

\begin{table*}[ht!]
    \centering 
    \label{packaging requirements}
    \caption{Major needs in the US advanced packaging ecosystem are required to be addressed for different application fields.}
    \centering
    \begin{tabular}{|>{\centering\arraybackslash}p{2.3cm}|>{\centering\arraybackslash}p{2.3cm}|>{\centering\arraybackslash}p{5.8cm}|>{\centering\arraybackslash}p{5.8cm}|}
    \hline
            \textbf{Major needs} & \textbf{Affected application fields} & \textbf{Concerns} & \textbf{Possible solutions} \\
    \hline
            \centering Ensuring domestic substrate supply & 
            \centering All
            &
            \begin{itemize}[noitemsep, left=0pt, topsep=0pt, partopsep=0pt, parsep=0pt, itemsep=0pt, after=\vspace{-\baselineskip}]
            \item Risk of potential bottleneck due to complete reliance on offshore IC and packaging substrate suppliers
            \end{itemize} &
            \begin{itemize}[noitemsep, left=0pt, topsep=0pt, partopsep=0pt, parsep=0pt, itemsep=0pt, after=\vspace{-\baselineskip}]
                \item Offering incentives to offshore substrate manufacturers to establish onshore substrate production to mitigate immediate substrate shortage risk
                \item Promoting US companies for establishing onshore substrate production
            \end{itemize}\\
    \hline
            \centering Decreasing offshore raw materials dependency: etching agent, photoresist & 
            \centering High-performance computing,5G communication, mobile, automotive
            &
            \begin{itemize}[noitemsep, left=0pt, topsep=0pt, partopsep=0pt, parsep=0pt, itemsep=0pt, after=\vspace{-\baselineskip}]
                \item Required for high-speed interconnect and RDL fabrication in 2.5D/3D packaging
            \end{itemize} &
            \begin{itemize}[noitemsep, left=0pt, topsep=0pt, partopsep=0pt, parsep=0pt, itemsep=0pt, after=\vspace{-\baselineskip}]
                \item Promoting domestic research for formulating the vital chemicals needed for the fabrication process
                \item Boosting the production of essential raw chemicals by the domestic chemical industry
            \end{itemize}\\
    \hline
              \centering Decreasing offshore semiconductor-grade pure chemical dependency: H2SO4, IPA & 
            \centering All
            &
            \begin{itemize}[noitemsep, left=0pt, topsep=0pt, partopsep=0pt, parsep=0pt, itemsep=0pt, after=\vspace{-\baselineskip}]
                \item Required in different stages of manufacturing and fabrication
                \item May affect future expansion and reshoring effort
            \end{itemize} &
            \begin{itemize}[noitemsep, left=0pt, topsep=0pt, partopsep=0pt, parsep=0pt, itemsep=0pt, after=\vspace{-\baselineskip}]
                \item Boosting the production of essential raw chemicals by domestic industry leaders such as DuPont, Dow Chemicals, etc.
            \end{itemize}\\
    \hline
            \centering Increasing onshore manufacturing capabilities & 
            \centering High-performance computing, 5G communication, mobile, automotive, Medical and Health, Internet of Things (IoT)
            &
            \begin{itemize}[noitemsep, left=0pt, topsep=0pt, partopsep=0pt, parsep=0pt, itemsep=0pt, after=\vspace{-\baselineskip}]
                \item The defense and aerospace sector exhibits self-reliance but faces high production costs and low output volumes
                \item Automotive sectors suffer from supply chain complexity due to having diverse packaging technology needs 
            \end{itemize} &
            \begin{itemize}[noitemsep, left=0pt, topsep=0pt, partopsep=0pt, parsep=0pt, itemsep=0pt, after=\vspace{-\baselineskip}]
                \item Incentives aimed at stimulating the establishment of OSATs
                \item Encouraging the integration of assembly and packaging capabilities within forthcoming fabrication facilities
            \end{itemize}\\
    \hline
            \centering Addressing future workforce demand & 
            \centering All
            &
            \begin{itemize}[noitemsep, left=0pt, topsep=0pt, partopsep=0pt, parsep=0pt, itemsep=0pt, after=\vspace{-\baselineskip}]
                \item Future expansion and reshoring efforts will be greatly jeopardized
            \end{itemize} &
            \begin{itemize}[noitemsep, left=0pt, topsep=0pt, partopsep=0pt, parsep=0pt, itemsep=0pt, after=\vspace{-\baselineskip}]
                \item Companies must offer competitive salaries to attract top talent
                \item Promoting packaging technology by integrating it into university curricula
            \end{itemize}\\
    \hline
       \centering Novel metrology methods to bridge the metrology gap (more stringent requirements) & 
            \centering All
            &
            \begin{itemize}[noitemsep, left=0pt, topsep=0pt, partopsep=0pt, parsep=0pt, itemsep=0pt, after=\vspace{-\baselineskip}]
               \item Conventional metrology techniques may fall short of meeting the increasingly stringent requirements of advanced packaging
               \item Risk of reverse engineering and counterfeiting will still persist even after reshoring
    
            \end{itemize} &
            \begin{itemize}[noitemsep, left=0pt, topsep=0pt, partopsep=0pt, parsep=0pt, itemsep=0pt, after=\vspace{-\baselineskip}]
                \item Promoting research for standardizing novel materials and material characterization, purity, and provenance 
                \item Developing provenance techniques for microelectronic components to improve trust and assurance
            \end{itemize}\\
    \hline
\end{tabular}
\end{table*}

Last but not least, it is also important to pay attention to how a US-based OSAT vendor market for advanced packaging could be structured in the future and devise potential solutions to mitigate major lackings affecting various applications sectors of advanced packaging (Table IV). In the current landscape, for instance, TMSC is the sole advanced packaging vendor for Apple. Although there are numerous benefits to developing a deep and rich relationship between a buyer and a vendor from a product development perspective, there are commendable supply chain risks that cannot be ignored as well. In the semiconductor industry, for instance, the 2000 fire in the Royal Philips Electronics wafer fabrication facility in Albuquerque has caused major disruptions for Nokia and Ericsson, both major buyers from Philips at the time\cite{noauthor_fire_nodate}.
The factory could not recover from the fire for about six weeks. While Nokia was able to recover from the loss by responding to the crisis and identifying alternative component providers for its mobile phone division, Ericsson could not move as quickly. Reporting an operating loss of \$200 million in the second quarter of 2000 in its mobile phone division, Ericsson could not fully recover from the fire. If Ericsson had more than one supplier at the time or had identified an alternative component supplier more quickly, the mobile phone market could have evolved quite differently. Hence, it is critically important to facilitate the development of deep and wide supplier bases for substrate, raw material and manufacturing supply providers, and OSAT vendors to ensure a secure US-based supply chain for advanced packaging operations. 

\subsection{Hardware Security Concerns}
Hardware security is a critical aspect of cybersecurity, and it encompasses a range of potential vulnerabilities and threats associated with the physical components of a system. Even if the supply chain is fully within the US and well-protected, the design and manufacturing processes can still introduce security vulnerabilities. Malicious actors like rogue employees might attempt to insert Hardware Trojans (HT), backdoors, or other malicious components during manufacturing stages or at any point in the supply chain. In addition, semiconductor manufacturing typically involves numerous subcontractors and suppliers. Each additional participant in the supply chain increases the potentiality for vulnerabilities and security gaps, regardless of location. Furthermore, despite having various countermeasures being developed against different types of HT\cite{varshney_shen_paradis_asadizanjani_2020}\cite{tehranipoor2023hardware}\cite{hoque_trust_2020}, the continuous evolution of sophisticated attacks and dependency on third-party IPs raises serious hardware security concerns \cite{farheen2021proof}\cite{gao2023iprobe}. 

Another major concern that always persists even if the US-based semiconductor supply chain is fully secured and becomes onshore is Reverse Engineering (RE). Chiplets involve extracting design details at the RTL level by reprocessing and imaging various device layers from fabricated ICs. Competing semiconductor design companies or adversarial foundries may use RE to gain a competitive edge and financial advantages\cite{meade_gate-level_2016}\cite{noauthor_survey_nodate}\cite{xi_physical_2022}\cite{true_terahertz_2021}. RE activities can result in revenue losses for the chiplet OEM and raise potential concerns about the reliability and trustworthiness of the reverse-engineered chiplet. Furthermore, similar to chips, reverse engineering chiplets is a time-consuming and labor-intensive process, which may lead adversaries to favor SiP reverse engineering due to its perceived benefits over chiplet reverse engineering. As the semiconductor packaging is getting increasingly complex, SiP RE will increasingly become new threats where adversaries will extract all the interconnects between the chiplets. Thus, it's important to ensure significant research to thwart SiP RE to deal with IP piracy and cloning problems\cite{rahman_physical_2018}\cite{biswas_backside_2022}\cite{biswas_emerging_2022}\cite{khan_security_2022}. 

To address these above hardware security concerns effectively, a multi-layered approach is essential. This approach includes secure design practices, thorough testing and verification processes, supply chain integrity checks, continuous monitoring, and implementing security best practices throughout the product lifecycle. Enhanced vulnerability management, including tracking materials and components across the product life cycle, is also essential for addressing these hardware security concerns. Additionally, developing and using trusted emerging techniques (AI and ML methods) across the entire semiconductor value chain is essential to enhance the security and provenance of microelectronic components and products across supply chains and increase trust and assurance.  

\subsection{Metrology Needs}
Securing the onshore US-based advanced packaging supply chain requires a robust metrology system. Metrology is the science of measurement and is crucial in ensuring the quality, reliability, and performance of advanced packaging technology. Semiconductor packaging is getting increasingly complex, and the yield rate is a real issue for ramping up advanced packaging manufacturing. Process control and yield management in semiconductor packaging rely heavily on metrology and inspection tools. Implementing metrology tools and techniques to monitor and control the various stages of the advanced packaging process is essential. This includes measurements during substrate fabrication, die-attach, wire bonding, encapsulation, and other packaging steps. Process control ensures consistency and adherence to specifications, reducing the risk of defects and ensuring the reliability of the packaged devices. These tools are crucial for ensuring the integrity of interconnects and monitoring various aspects of the bumping and bonding processes.  

Metrology focuses on parameters such as diameter, height, and co-planarity for bumping processes. Accurate dimensional measurements are critical in advanced packaging to ensure the precise placement and alignment of microelectronic components. This includes measuring the dimensions of interconnects, micro-bumps, under-bump metallization (UBM), and other critical features. Smaller pitches require tighter control over bump diameter and height, and as bump height decreases, the acceptable window of co-planarity becomes narrower. Meeting the increasingly stringent requirements for semiconductor materials purity, physical properties, and provenance of materials used in advanced packaging across a diverse supply chain is also necessary. To satisfy these stringent requirements, new measurements and standards are crucial. These advancements in metrology will help ensure the quality, reliability, and security of the supply chain. Furthermore, developing measurement technologies, properties data, and standards focused on defect and contaminant identification to support uniform materials quality and traceability across the supply chain should also be a significant strategic focus.  

In addition, breakthroughs in physical and computational metrology are crucial for advancing manufacturing techniques for future-generation devices due to their need for precision, accuracy, data integration, digital Twining, and more. Advancements in physical and computational metrology are indispensable for the successful manufacturing of future-generation devices. These breakthroughs can lead to improved precision, enhanced real-time monitoring, and the ability to handle the complexity and challenges associated with cutting-edge technologies. Ensuring that critical metrology advances keep up with cutting-edge and future microelectronics and semiconductor manufacturing, advancement in the physical and computational metrology tools adaptable to next-generation manufacturing of advanced complex, integrated technologies and systems is necessary. 

Indeed, as microelectronics packaging continues to evolve, new metrology techniques are crucial to integrating sophisticated components and novel materials. In this context, Metrology refers to the science of measurement and characterization of the tiny components and structures that comprise advanced microelectronics packages. The traditional metrology methods may not be sufficient to meet the demands of emerging technologies and materials used in microelectronics packaging due to several factors, including: i) shrinking feature sizes, ii) multilayer and multi-material stacks, iii) physical properties for films, surfaces, buried features, and interfaces iv) Methods for integrating chiplets, dialects, Systems on a Chip (SoCs), and memories into packages. Provide enabling metrology that spans multiple length scales and physical properties and supports accelerating advanced packaging concepts for future-generation microelectronics. Developing metrology for the complex integration of sophisticated components and new materials is necessary to support a solid domestic advanced microelectronics packaging industry. Overall, investment in both process and metrology tools is essential to achieve the required quality and reliability, especially as the complexity of semiconductor packaging increases\cite{noauthor_metrology_nodate}.

\subsection{Workforce Development Needs}
Despite leading the global market in chip IP design and equipment manufacturing, the US faces a major setback in strengthening domestic manufacturing due to a significant talent shortage, spanning technicians to design engineers. This shortage has critical implications, as it hampers the country's ability to maintain a secure manufacturing capacity for semiconductor chips, which are vital for defense systems, automation, and quantum computing. The industry's workforce development encounters key challenges, such as a lack of student interest in hardware electronics, an outdated curriculum that neglects modern semiconductor techniques, talent retention issues, and an aging faculty and infrastructure. Addressing these obstacles is crucial for fostering the industry's future growth and innovation. To ensure onshore advanced packaging capabilities in the US, addressing workforce development is essential to fully secure the semiconductor supply chain \cite{noauthor_fueling_nodate}. 
Decades of outsourcing have led to a lack of familiarity and prestige in semiconductor careers compared to software jobs in big tech companies. The industry's poor perception and limited awareness of its dynamic nature exacerbate the talent shortage. Although the US has access to a vast talent pool through international STEM students, most engineering Master's and Ph.D. students are non-US citizens. Onerous visa restrictions also contribute to low retention of international students, leading to a loss of valuable talent.

To remain competitive in the talent war, companies are exploring strategies like reskilling, automation, and expanding the talent pipeline. However, these efforts require significant time and capital and may not be sufficient to offset the projected shortage of engineers and technicians in the industry. Additionally, the complex fabrication process in fast-paced fabs requires a lengthy learning curve for new or retrained employees. There is also an income discrepancy between design engineers and quality engineers/manufacturing technicians, with design engineers enjoying better working conditions and higher pay. This makes re-skilling employees for higher-paying desk jobs challenging. Another issue is the OSATs run on razor-thin margins. Attractive remuneration should be offered to attract talented engineers in the packaging industry. To ensure a thriving future for the semiconductor and packaging technology sectors, we must invest in and encourage the younger generation at the national level. Demonstrating the benefits of automation throughout the manufacturing process is essential in capturing the interest of young individuals. By showcasing the flexibility and remote-control capabilities of modern equipment, we can make these fields more appealing to emerging talent. In conclusion, the semiconductor industry in the US must address these talent-related issues to meet the advanced packaging demand and sustain future growth\cite{noauthor_addressing_nodate}\cite{noauthor_reshoring_nodate}\cite{rizi2023talent_2}.

\section{Conclusion}
Advanced packaging, along with heterogeneous integration, is one of the key enablers to advance Moore's Law. Although, due to the CHIPS act, there is significant attention by the government to ramp up on-shore chip fabrication capabilities, it's equally important to give the same attention to advanced packaging manufacturing capabilities, too. Understanding the current US-based advanced packaging supply chain is essential to identify the bottleneck and future direction to fully secure the semiconductor supply chain and develop onshore advanced packaging manufacturing capabilities. Although the US has all the advanced packaging manufacturing capabilities, significant investment is required to ramp up the onshore production. Besides ramping up the manufacturing capabilities, securing the essential raw material supplies for advanced packaging manufacturing processes, such as substrates, molding compounds, lead-frames, etching agents, photoresists, and semiconductor-grade ultra-purified raw chemicals, is essential to avoid future bottlenecks. The absence of packaging substrate material production in the US, which may affect all application fields of advanced packaging, needs to be solved. The US advanced packaging supply chain can be secured by solving all of these weak links in the supply chain.

\section*{Acknowledgement}
The authors wish to express their sincere gratitude to Dr. Nelson Hastings of the National Institute of Standards and Technology (NIST) for his valuable feedback and suggestions in this work. 

{
\printbibliography
}
\end{document}